\documentclass{PoS}
\usepackage{amsmath}
\usepackage{amsfonts,amssymb}

\title{Spectral Action
from Anomalies}

\ShortTitle{Spectral Action
from Anomalies}

\author{A.A.~Andrianov$^{abc}$, M.A.~Kurkov$^a$~ \speaker{Fedele Lizzi}\,$^{bcde}$,~ \\
        $^{a}$ V.A. Fock Department of Theoretical Physics, Sankt-Petersburg State University,
198504 St. Petersburg, Russia\\
$^{b}$\it High Energy Physics Group, Dept. Estructura i Constituents
de la Mat\`eria, \\Universitat de Barcelona, Diagonal 647, 08028
Barcelona, Catalonia, Spain \\
$^{c}$
Institut de Ci\`encies del Cosmos, UB, Barcelona\\
 $^{d}$ Dipartimento di Scienze Fisiche, Universit\`{a} di
Napoli {\sl Federico II}~\\ $^e$ INFN, Sezione di Napoli\\
Monte S.~Angelo, Via Cintia, 80126 Napoli, Italy\\
        E-mail: \email{andrianov@bo.infn.it,  katok86@mail.ru, fedele.lizzi@na.infn.it\\
        Report \# DSF/1/2011, ICCUB-11-059, SPbU-IP-11-01}}

\abstract{ Starting from a theory of fermions moving in a fixed
gauge and gravitational background we implement the scale
invariance of the theory. Upon quantization the theory is
anomalous but the anomaly can be cancelled by the addition of
another term to the action. This term comes out to be basically
the Chamseddine Connes spectral action introduced in the context
of noncommutative geometry. An alternative realization of the
dilaton may involve a collective scalar mode of all fermions
accumulated in a {scale-noninvariant} dilaton action. The entire
spectral action describes gauge and Higgs fields coupled with
gravity. Here this action is coupled with a dilaton and we discuss
how it relates to the transition from the radiation to the
electroweak broken phase via condensation of Higgs fields.}

\FullConference{Corfu Summer Institute on Elementary Particles and Physics - Workshop on Non
            Commutative Field Theory and Gravity,\\
        September 8-12, 2010\\
        Corfu Greece}

\newcommand{\e}{{\rm e}}
\newcommand{\ii}{{\rm i}}
\newcommand{\dd}{{\rm d}}
\newcommand{\eqn}[1]{(\ref{#1})}
\newcommand{\be}{\begin{equation}}
\newcommand{\ee}{\end{equation}}
\newcommand{\beq}{\begin{equation}}
\newcommand{\eeq}{\end{equation}}
\newcommand{\bea}{\begin{eqnarray}}
\newcommand{\eea}{\end{eqnarray}}
\newcommand{\ba}{\begin{eqnarray}}
\newcommand{\ea}{\end{eqnarray}}

\def\bra#1{\left\langle #1\right|}
\def\ket#1{\left| #1\right\rangle}

\def\ketbra#1#2{\left| #1\right\rangle\left\langle #2\right|}

\newcommand{\tr}[1]{\:{\rm tr}\,#1}
\newcommand{\Tr}[1]{\:{\rm Tr}\,#1}
\def\one{\mbox{1 \kern-.59em {\rm l}}}
\newcommand{\del}{\partial}

\newcommand{\complex}{{\mathbb C}} 

\begin{document}

\section{Introduction}
The aim of this note is to show how, starting from a theory of
fermions coupled to a gauge and gravitational background, it is
possible to have the full bosonic action emerged. We do this using
the spectral properties of the (generalized) Dirac
operator, and in this respect this work can be seen in the
framework of Connes and collaborators approach to the standard
model~\cite{Connesbook, ConnesLott, SpectralAction, AC2M2}, as well
as of Sakharov induced gravity~\cite{Sakharov} (for a modern review
see~\cite{Visser}). Our starting point is a theory of fermions
moving in a fixed background of  gauge and scalar fields and a curved
(Euclidean) spacetime. We focus on the scale invariance of the
theory at the classical level. To quantize it we employ the spectral
regularization, based on the cutoff of the eigenvalues of the Dirac
operator. The regularization however is not preserved at the quantum
level, and a scale anomaly is developed. There are two alternative ways to deal with quantum anomalies in local transformations \cite{aano}. On one hand the scale invariance can be
restored by changing the measure in the path integral. This is
tantamount to the addition to the fermionic action of another
contribution, which renders the bosonic background interacting to the dilaton field.  An alternative realization of the dilaton may involve a collective scalar mode of all fermions accumulated in a {scale-noninvariant} dilaton action. Accordingly the spectral action arises as a part of the fermion effective action divided into the scale non-invariant and scale invariant parts. It
turns out that in both cases the resulting action is a modification of
Chamseddine-Connes spectral action but with opposite signs. The latter is a purely spectral
function of the gauged Dirac operator (and of a cutoff) which
describes a gauge theory coupled with gravity, and in the presence
of the Higgs mechanism.

The invariance by (local) scale transformation introduces in the
theory another field whereas the collective dilaton mode of fermions appears after their partial bosonization as a consequence of scale non-invariance. The dilaton effective potential has been
calculated by us and in the last section we discuss how it relates
to the transition from the radiation phase with zero v.e.v. of Higgs fields and massless particles
to the electroweak broken phase via condensation of Higgs fields. It is proven that only the second way to interpret the spectral action with collective field of dilaton can provide the above mentioned phase transition with EW symmetry breaking during Universe evolution.

The first six Sections of these proceedings will mostly follow
reference~\cite{anlizzi}, although with a somewhat different point
of view.  The material in Sec.7 has not been previously
published.

\section{Fermions in a Fixed Background}

Our starting point is a theory in which we have some matter fields,
represented by fermions transforming under some (reducible)
representation a gauge group, such as the standard model group
$SU(3)\times SU(2)\times U(1)$. We need not specify the group for
the moment. The fermions will be spinors belonging to some Hilbert
space $\cal H$ which we assume to be ``chiral'', i.e.\ split into a
left and a right:
\be
{\cal H}={\cal H}_L\oplus{\cal H}_R
\ee
A generic matter field  will therefore be a spinor
\be
\Psi=\begin{pmatrix}\Psi_L\\
\Psi_R\end{pmatrix}
\ee
and in this representation the chirality operator, which we call
$\gamma$ is a two by two diagonal matrix with plus and minus one
eigenvalues. The two components are spinors themselves and we are
not indicating the gauge indices, nor the flavor indices. We will
assume that the fermions come in a number of identical (apart from
the mass) generations.

The fermions are given a dynamics  coupling them to a background
field. This coupling is performed by a classical action which we
schematically write as
\be
S_F=\bra{\Psi}D\ket{\Psi} \label{fermionicaction}
\ee
where
\be D=D_0+A
\ee
is an operator on $\cal H$ which will call always the Dirac
operator, although the formalism we are building is more general and
there may be ``Dirac operators'' which do not resemble at all the
one introduced for the Dirac equation.

The Dirac operator, acting of spinors is again a matrix and we have
split into a ``free and gravitational'' part and a ``gauge
coupling'' part. We will see in a moment the reason for this (rather
inaccurate) terminology.

We start from
\be
D_0=\begin{pmatrix}\gamma^\mu\del_\mu
& M\\ M^\dagger & \gamma^\mu\del_\mu
\end{pmatrix}
\ee
Where $M$  contains all masses (and mixings) of the fermions and the
$\gamma$ are those relative to a possibly curved spacetime. In this
case the fermions are coupled to the gravitational fixed background
given by the metric
\be
g^{\mu\nu}=\frac12 \{\gamma^\mu,\gamma^\nu\}
\ee

The matrix $A$ represents instead a fixed gauge background, and the
interaction of the spinors with it. We emphasize that at this stage
we are just describing the classical dynamics of fermions in a fixed
background. We are deliberately  vague as to the detail of the model
at this stage, not discussing important elements of the theory, like
chirality or charge conjugation. The scheme presented here is
largely independent on the details of the model. In particular it
applies to the standard model, especially in the approach based on
noncommutative geometry introduced by Connes and briefly discussed
below.

\section{Fields, Hilbert Spaces, Dirac Operators and the
(Non)commutative Geometry of Spacetime}

We have introduced a (Euclidean) spacetime. And therefore
implicitly the algebra $\cal A$ of complex valued continuous
functions of this space time. There is in fact a one-to one
correspondence between (topological Hausdorff) spaces and
commutative $C^*$-algebras, i.e.\ associative normed algebras with
an involution and a norm satisfying certain properties. This is
the content of the Gelfand-Naimark theorem~\cite{FellDoran,
Ticos}, which describes the topology of space in terms of the
algebras. In physicists terms we may say the properties of a space
are encoded in the continuous fields defined on them. This
concept, and its generalization to noncommutative algebras is one
of the starting points of Connes noncommutative geometry
programme~\cite{Connesbook}. The programme aims at the
transcription of the usual concepts of differential geometry in
algebraic terms and a key role of this programme is played by a
\emph{spectral triple}, which is composed by an algebra acting as
operators on a Hilbert space and a (generalized) Dirac operator.
In our case we have these ingredients, but we have to consider
instead of the algebra of continuous complex valued function,
matrix valued functions. The underlying space in this case is
still the ordinary spacetime, technically the algebra is ``Morita
equivalent'' to the commutative algebra, but the formalism is
built in a general way so to be easily generalizable to the truly
noncommutative case, when the underlying space may not be an
ordinary geometry.

The spectral triple contains the information on the geometry of
spacetime. The algebra as we said is dual to the topology, and the
Dirac operator enables the translation of the metric and
differential structure of spaces in an algebraic form. There is no
room in these proceedings to describe this programme, and we refer
to the literature for details~\cite{Connesbook, Landibook, Ticos,
Madore}.

Within this general programme a key role is played by Connes'
approach to the standard model. This  is the attempt to understand
which kind of (noncommutative) geometry gives rise to the standard
model of elementary particles coupled with gravity. The roots of
this approach is to have the  Higgs appear naturally as the
``vector'' boson of the internal noncommutative degrees of
freedom~\cite{Madoreearly, D-VKM, ConnesLott}. The most complete
formulation of this approach is given by the \emph{spectral action},
which in its most recent form is presented in~\cite{AC2M2}.

The fermionic part of this action is~\eqn{fermionicaction}, while
the bosonic part is basically the regularized sum of the eigenvalues
of the Dirac operator. We will see how this action can fe inferred
(with some little modifications) from the fermionic action $S_F$ and
the need to preserve scale invariance.

\section{Scale invariance of the Fermionic Action}

So far we are in the presence of a classical theory of matter fields
moving in a fixed background. The objects involved in the writing of
the action have physical dimensions. Introduce a scale necessary for
measurements, for example an unit of length $\ell$, then it is
possible to measure volumes as $\ell^{-4}$, masses\footnote{We take
the speed of light $c=1$.} and the Dirac operator in general as
$\ell^{-1}$ and so on.

The classical action is invariant under a change of this scale,
after all is amounts to just a change of units of measurement.
Recall that we have not yet introduced $\hbar$. In principle this
change of scale could also be local, and this would be H.~Weyl
original ``gauge'' theory. We therefore have a scale transformation
symmetry:
\bea
x^\mu&\to&\e^\phi x^\mu \nonumber\\ \psi&\to& \e^{-\frac32\phi}
\psi\nonumber\\ D&\to& \e^{-\frac12\phi}D\e^{-\frac12\phi}
\label{scaleinvariance}
\eea
where $\phi$ is a real parameter which for the moment we take to be
constant. Note that since the rescaling involves also the matrix
part of $D$, we must also rescale the masses of the fermions. In the
absence of a dimensional scale, is an exact symmetry of the
classical theory.

We now proceed to quantize the theory. It can be proven \cite{Fujikawabook} that if the
classical theory is invariant, the measure in the quantum path
integral is not. We have an anomaly: a classical theory is invariant
against a symmetry transformation, but the quantum theory, due to unavoidable
regularization, does not possess this symmetry anymore. If also the
quantum theory is required to be symmetric then the symmetry can be
restored by the addition of extra terms in the action. A textbook
introduction to anomalies can be found in \cite{Fujikawabook}. The
notion of scale anomaly is attached to the dilatation  of both
coordinates,  fields and mass-like parameters according to their
dimensionalities, Eq.~\eqn{scaleinvariance}. Evidently, in the
absence of UV divergences, there is no scale anomaly which therefore
can be correlated to rescaling of a cutoff in the theory. In the case when
the dilatation is  not constant,  $\phi$ becomes a quantum field
called the \emph{dilaton}. The dilaton of this kind has been investigated in the
context of the spectral action in~\cite{ChamseddineConnesscale}.

We remark that there is also an alternative realization of the dilaton as a collective scalar mode of all fermions accumulated in a {scale-noninvariant} dilaton action. The corresponding spectral action has an opposite sign and will be discussed later on.

In both approaches we start from the partition function
\be
Z(D)=\int [\dd\psi] [\dd\bar\psi] e^{-S_\psi} =\det\left(\frac{D}{\mu}\right)
\ee
where a normalization scale $\mu$ is introduced and the last equality is formal because the expression is
divergent and needs  regularizing. The writing of the fermionic
action in this form (as a Pfaffian) is instrumental in the solution
of the fermion doubling problem in Connes approach to the standard
model~\cite{LMMS, G-BIS, AC2M2}. In fact we need in principle two
regulators:

\begin{itemize}
\item {$\mu$ which may be treated as an infrared cutoff  when having a discrete
    spectrum;}
\item {an ultraviolet cutoff $\Lambda$ in order to tame the
    short distance infinities.}
\end{itemize}

We will regularize the theory in the ultraviolet using a procedure
introduced by one of us, Bonora and Gamboa-Saravi
in~\cite{AndrianovBonoraGamboa, AndrianovBonora1, AndrianovBonora2} but leaving a room for a normalization scale $\mu$.
The energy cutoff is enforced by considering only the first $N$
eigenvalues of $D$. Consider the projector
\be
P_N=\sum_{n=1}^N \ketbra{\lambda_n}{\lambda_n};\quad N=\max n \
\mbox{such that} \ \lambda_n\leq \Lambda \label{cuteigenvalues}
\ee
where $\lambda_n$ are the eigenvalues of $D$ in increasing order
(repeated according to possible multiplicities), $\ket{\lambda_n}$ a
corresponding orthonormal basis, and  the integer $N$ is a function
of the cutoff. This means that we are effectively using the
$N^{\mathrm{th}}$ eigenvalue as cutoff. Therefore this number and
the corresponding spectral density depends on coefficient functions
of the Dirac operator, $N=N(D)$.

In the framework of noncommutative geometry this is the most natural
cutoff procedure, although it was introduced before the introduction
of the standard model in noncommutative geometry. It makes no
reference in principle to the underlying structure of spacetime, and
it is based purely on spectral data, thus is perfectly adequate to
Connes' programme. This form of regularization could be also used
for field theory which cannot be described on an ordinary spacetime,
as long as there is a Dirac operator, or generically a wave
operator, with a discrete spectrum.

We define the regularized partition function\footnote{Although $P_N$
commutes with $D$ we prefer to use a more symmetric notation.}
\ba
Z_\mu(D)&=&\prod_{n=1}^N\frac{\lambda_n}{\mu}
=\det\left(\one-P_N+P_N\frac{D}{\mu}P_N\right)\nonumber\\
&=& \det\left(\one-P_N+P_N\frac{D}{\Lambda}P_N\right)
\det\left(\one-P_N+\frac{\Lambda}{\mu}P_N\right)= Z_\Lambda(D) \det\left(\one-P_N+\frac{\Lambda}{\mu}P_N\right).
\ea
In this way we can define the fermionic action in an intrinsic way.

The regularized partition function $Z_\Lambda$ has a well defined
meaning. Expressing $\psi$ and $\bar\psi$ as
\bea
\psi=\sum_{n=1}^\infty a_n\ket{\lambda_n};\qquad
\bar\psi=\sum_{n=1}^\infty b_n\ket{\lambda_n}
\eea
with $a_n$ and $b_n$ anticommuting (Grassman) quantities. Then
$Z_\Lambda$ becomes (performing the integration over Grassman
variables for the last step)
\be
Z_\Lambda(D)=\int\prod_{n=1}^N \dfrac{\dd a_n \dd b_n}{\Lambda} \e^{-\sum_{n=1}^N b_n
\lambda_n
 a_n}=\det\left(D_N\right)
\ee
where we defined
\be
D_N=1-P_N+P_N\frac{D}{\Lambda}P_N .
\ee
In the basis in which $D/\Lambda$ is diagonal it corresponds to set
to $\Lambda$ all eigenvalues larger than $\Lambda$. Note that $D_N$
is dimensionless and depends on $\Lambda$ both explicitly and
intrinsically via the dependence of $N$ and $P_N$.

It is possible to give an explicit functional expression to the
projector in terms of the cutoff:
\be
P_N=\Theta\left(1-\frac{D^2}{\Lambda^2}\right)=\int\limits_{-\infty}^{\infty}\dd\alpha\,\frac1{2\pi\ii(\alpha-
\ii\epsilon)} \e^{\ii\alpha\big(1-\frac{D^2}{\Lambda^2}\big)}
\ee
where $\Theta$ is the Heaviside step function. This integral is well
defined for a compactified space volume and therefore in the
presence of the infrared cutoff which can be identified with $\mu$. Actually $N$ depends also on
the infrared cutoff, and the number of dimensions. It goes as
$\sim\left(\frac{\Lambda}{\mu}\right)^d$.

\section{Cancellation of the Anomaly and the Bosonic Action}

Let's perform the first scenario and restore scale invariance. The
action $S_F$ is invariant under~\eqn{scaleinvariance} but the
partition function is not, the reason for this is the fact that
the regularization procedure is not scale invariant. The
cancellation of the anomaly then proceeds via a change of measure,
which is equivalent to the addition of another term to the action.
This term compensates the change in the measure due to the
regularization, but being in an exponential form, can also be seen
as another addition to the action, so that the final partition
function is invariant. This calculation has been performed
in~\cite{AANN} in the QCD context, and applied to gravity
in~\cite{NovozhilovVassilevich}.

Let us see in a very heuristic way, with $\phi$ constant, why the
effective action $S_{\mathrm{eff}}$ is nothing but the spectral
action with the function $\chi$ being a sharp cutoff. In this case
$N$ is just a number of eigenvalues smaller that $\Lambda$, and
thereby
\be
\Tr\chi\left(\frac{D^2}{\Lambda^2}\right)=\Tr\Theta\left(1-\frac{D^2}{\Lambda^2}\right)=\Tr P_N= N(\Lambda,\ D).
\ee
It can be written in the latter form provided that we take into
account the functional dependence $ N=N(\Lambda,\ D)$. It is worth
recalling again that the integer $N$ depends on the cutoff
$\Lambda$, on the Dirac operator $D$ and also on the function $\chi$
which we have chosen to be a sharp cutoff.

Then the compensating term -- the effective action, will be defined
by
\be
{Z_{\mathrm{inv}}}_\mu(D)=Z_\mu(D) \int\dd\phi\,
\e^{-S_{\mathrm{anom}}}
\ee
where the effective action will be depending on $N$, and hence the
cutoff $\Lambda$, and on $\phi$. Define
\be
Z_{\mathrm{inv}\mu}(D)=\int\dd\phi Z_\mu(\e^{-\frac12\phi}
D\e^{-\frac12\phi})= \int\dd\phi Z_\mu(
D_\phi);\qquad D_\phi\equiv \e^{-\frac12\phi}
D\e^{-\frac12\phi} ,
\ee
then
\be
S_{\mathrm{anom}}= \log Z_\mu(D)Z_\mu^{-1}( D_\phi) \label{sanom1}
\ee

Let us assign
\be
Z_t=Z_\mu(D_{t\phi})
\ee
therefore $Z_0=Z_\mu(D)$ and
\be
{Z_{\mathrm{inv}}}_\mu(D)Z^{-1}_\mu(D)=\int\dd\phi \frac{Z_1}{Z_0}
\ee
and hence
\be
S_{\mathrm{anom}}=-\int_0^1\dd t \del_t \log Z_t =-\int_0^1\dd t
\frac{\del_t Z_t}{Z_t}
\ee
We have the following relation that can easily proven
\bea
\del_t Z_t=\del_t\det \left(\frac{D_{t\phi}}{\mu}\right)_N &=&\phi
Z_t  \left(-1 +\Lambda^2 \log\frac{\Lambda^2}{\mu^2}
\partial_{\Lambda^2}\right) \tr P_N ,
\eea
and therefore
\bea
S_{\mathrm{anom}}&=&\int_0^1\dd t\, \phi \left(1 -\Lambda^2
\log\frac{\Lambda^2}{\mu^2}
\partial_{\Lambda^2}\right)\Tr\Theta\left(1-\frac{D_{t\phi}^2}{\Lambda^2}\right)\nonumber\\
&=& \int_0^1\dd t\, \phi \left(1 -\Lambda^2
\log\frac{\Lambda^2}{\mu^2} \partial_{\Lambda^2}\right)
N(\Lambda,\ D_{t\phi}).\ \label{Sanomal}
\eea

\section{The Spectral Action}

For $\phi=0$ this is basically the Chamseddine-Connes Spectral
Action introduced in~\cite{SpectralAction} together with the
fermionic action~\eqn{fermionicaction}. More precisely the bosonic
part of the spectral action is
\be
\Tr \chi\left(\frac{D^2}{\Lambda^2}\right)
\ee
where $\chi$ is a generic cutoff function, which in our case is a
sharp cutoff at energy $\Lambda$,
\be
\chi(x)=\left\{\begin{array}{cc}0~ & x<0\\ 1~ & x\in[0,1]\\ 0~ & x>1
\end{array}\right. \label{sharpcutoff}
\ee
consequence of the sharp cutoff on the eigenvalues used
in~\eqn{cuteigenvalues}. The bosonic spectral action so introduced
is always finite by its nature, it is purely spectral and it
depends on the cutoff $\Lambda$. In the original work of
Chamseddine and Connes the bosonic and fermion parts of the action
were treated differently. The fermionic action on the contrary is
divergent, and will require renormalization. It is formulated as
an usual integral. In the philosophy of noncommutative geometry
usual integrals can be interpreted as a regularized trace, the
Dixmier trace:
\be
\int \dd x f=\Tr_\omega |D|^{-4} f
\ee
where the Dixmier trace of an operator $O$ with eigenvalues $o_n$
(ordered in decreasing order, repeated in case of degeneracy) is:
\be
\Tr_\omega O=\lim_{N\to\infty}\frac{1}{\log N} \sum_{n=0}^No_n
\ee
The integral/Dixmier trace has however to be regularized. We have
seen as the cancellation of the anomaly brings the two actions on
the same footing, albeit with a modification of the bosonic part.
It must be mentioned that already in~\cite{Sitarz} the two actions
are ``unified'' in the bosonic action with the addition of the
projection on the fermionic field to the covariant Dirac operator.
This reproduces the full spectral action with some additional non
linear terms for the fermions, which could have to do with
fermionic masses.

To obtain the standard model take as algebra the product of the
algebra of functions on spacetime times a {finite dimensional}
matrix algebra
\be \mathcal A =C({\mathbb
R}^4)\otimes{\mathcal A}_F
\ee
Likewise the Hilbert space is the product of fermions times a finite
dimensional space which contains all matter degrees of freedom, and
also the Dirac operator contains a continuous part and a discrete
one
\be
\mathcal H ={\mathrm{Sp}({\mathbb R}^4)\otimes{\mathcal H}_F}
\ee
and the Dirac operator
\be
D_0=\gamma^\mu\del_\mu\otimes\mathbb I + \gamma\otimes D_F
\ee
In its most recent form  due to Chamseddine, Connes and
Marcolli~\cite{AC2M2} a crucial role is played by the mathematical
requirements that the noncommutative algebra satisfies the
requirements to be a manifold. Then the internal algebra, is almost
uniquely derived to be
\be
{\mathcal A}_F=\complex\oplus{\mathbb H}\oplus M_3(\complex)
\ee
Then the  bosonic spectral action can be evaluated at one loop
using standard heath kernel techniques~\cite{Vassilevich:2003xt}
and the final result gives the full action of the standard model
coupled with gravity. We restrain from writing it since it takes
more than one page in the original paper~\cite{AC2M2}. In the
process however one does not need to input the mass of the Higgs,
which comes out as a prediction. Its value comes out to be $\sim
170 \mathrm{GeV}$. A small value experimentally disfavoured. It
must be said however that the present form of the model needs
unification of the three coupling constant at a single energy
point (given by $\Lambda$). The model also contains nonstandard
gravitational terms (quadratic in the curvature), which are
currently being investigated for their cosmological
consequences~\cite{NelsonSakellariadu, MarcolliPierpaoli}.

Technically the canonical bosonic spectral action is a sum of
residues, and can be expanded in a power series in terms of
$\Lambda^{-1}$ as
$$
S_B=\sum_n f_n\, a_n(D^2/\Lambda^2)
$$
where the $f_n$ are the momenta of $\chi$
\begin{eqnarray*}
f_0&=&\int_0^\infty \dd x\, x  \chi(x)\nonumber\\
f_2&=&\int_0^\infty \dd x\,   \chi(x)\nonumber\\
f_{2n+4}&=&(-1)^n \del^n_x \chi(x)\bigg|_{x=0} \ \ n\geq 0
\label{fcoeff}
\end{eqnarray*}
the $a_n$ are the Seeley-de Witt coefficients which vanish for $n$
odd. For $D^2$ of the form
$$
D^2=g^{\mu\nu}\del_\mu\del_\nu\one+\alpha^\mu\del_\mu+\beta
$$
defining
\begin{eqnarray*}
\omega_\mu&=&\frac12 g_{\mu\nu}\left(\alpha^\nu+g^{\sigma\rho} \Gamma^\nu_{\sigma\rho}\one\right)\nonumber\\
\Omega_{\mu\nu}&=&\del_\mu\omega_\nu-\del_\nu\omega_\mu+[\omega_\mu,\omega_\nu]\nonumber\\
E&=&\beta-g^{\mu\nu}\left(\del_\mu\omega_\nu+\omega_\mu\omega_\nu-\Gamma^\rho_{\mu\nu}\omega_\rho\right)
\end{eqnarray*}
then
\begin{eqnarray*}
a_0&=&\frac{\Lambda^4}{16\pi^2}\int\dd x^4 \sqrt{g}
\tr\one_F\nonumber\\
a_2&=&\frac{\Lambda^2}{16\pi^2}\int\dd x^4 \sqrt{g}
\tr\left(-\frac R6+E\right)\nonumber\\
a_4&=&\frac{1}{16\pi^2}\frac{1}{360}\int\dd x^4 \sqrt{g}
\tr(-12\nabla^\mu\nabla_\mu R +5R^2-2R_{\mu\nu}R^{\mu\nu}\nonumber\\
&&+2R_{\mu\nu\sigma\rho}R^{\mu\nu\sigma\rho}-60RE+180E^2+60\nabla^\mu\nabla_\mu
E+30\Omega_{\mu\nu}\Omega^{\mu\nu}) \label{spectralcoeff}
\end{eqnarray*}
$\tr$ is the trace over the inner indices of the finite algebra
$\mathcal A_F$ and in $\Omega$ and $E$ are contained the gauge
degrees of freedom including the gauge stress energy tensors and the
Higgs, which is given by the inner fluctuations of $D$.

In our case
for $\phi$ constant, after performing the integration we find
\bea
S_{\mathrm{anom}}&=&\int_0^\phi \dd t' \sum_n \e^{(4-n)t'}\left(1 -\Lambda^2 \log\frac{\Lambda^2}{\mu^2} \partial_{\Lambda^2}\right) a_n f_n\nonumber\\&=&
\frac{1}{8} (e^{4\phi}-1) a_0\left(1 - 2\Lambda^2 \log\frac{\Lambda^2}{\mu^2} \right) + \frac{1}{2} (e^{2\phi}-1) a_2 \left(1 - \Lambda^2 \log\frac{\Lambda^2}{\mu^2} \right)+ \phi
a_4 .\label{Sanomaaction}
\eea
There are just  some numerical corrections to the first two
Seeley-de Witt coefficients due to the integration in $t\phi$ and a choice of normalization scale $\mu$.

We notice that the alternative way of treatment of the scale degree of freedom as a collective field leads to precisely the opposite sign of the dilaton action $S_{anom} \to S_{coll} = - S_{anom}$. Indeed the bosonization in scale variable can be represented as,
\be
Z_\mu(D)=Z_{\mathrm{inv},\mu}(D) \int\dd\phi\,
\e^{-S_{\mathrm{coll}}};\quad Z_{\mathrm{inv}\mu}(D)= \int\dd\phi Z^{-1}_\mu(
D_\phi)
\ee
then
\be
S_{\mathrm{coll}}= \log Z^{-1}_\mu(D)Z_\mu( D_\phi) = - S_{anom},
\ee
(cf. to \eqref{sanom1}) .
\section{The Dilaton and the effective potential}
The full analysis of the model coupled with a dynamical dilaton is
under way and will be published elsewhere. Nevertheless it is
already possible to say something on the interplay between the
dilaton and the Higgs, and in particular the effective potential.
This can be used to characterize cosmic evolution right after
inflation starts. In particular, it may open the ways to describe
the transition from the radiation phase with massless particles to
the EW symmetry breaking phase with spontaneous mass generation due
to condensation of Higgs fields.

\subsection{Mass generation from Higgs-dilaton potential during
cosmic evolution} We will consider in the following only the
potential terms relative to the complex Higgs doublet $H$ and the
dilaton $\phi$. The quadratic term of the Higgs potential comes from
the $a_2$ term of \eqn{Sanomaaction}, while the quartic one comes
from the $a_4$ one. In this way we can derive the form effective
Higgs-dilaton potential. To focus on this goal we reduce the joint
effective Higgs-dilaton (HD) potential including only the real
scalar component $H$ of the Higgs doublet $ (H_1, H_2) \to (0, H) $
subject to condensation. After performing renormalization the
general form of the HD potential is expected to be,
\be
V = V_0 + A e^{4\phi} + BH^2e^{2\phi} - CH^4(\phi+ \phi_0) + EH^2,
\label{1}
\ee
where depending on the normalization scale $\mu$ of fermion
effective action compared with the cutoff $\Lambda$ one can get any
sign of the coefficients $A(\Lambda,\mu), B(\Lambda,\mu)\gtrless 0
$. Evidently the constant $\phi_0$ can be eliminated by shifting the
field $\phi \to \phi-\phi_0$ and rescaling the constants $A,B$. Thus
in general both signs and modules  of these constants $A,B$ don't
have any a priori values. As to the constant $C$, if the dilaton
serves for restoration of conformal symmetry as an independent field
then the conformal anomaly coefficient $C < 0$ (see~\cite{anlizzi}).
On the other hand,   a composite dilaton made of fermions
\cite{AANN} has an anomalous part of the potential of the opposite
sign with $C > 0$. Therefore the sign of $C$ characterizes the
nature of the dilaton field: elementary or composite one. In this
Section we are interested in evolution of fields $\phi, H$ and
correspondingly neglect the additional cosmological constant $V_0$.
Thus for our purpose the potential has four arbitrary parameters
$A,B,C,E$ and $\phi_0 = 0$.

We would like to apply the HD potential for description of cosmic
evolution and select out of the acceptable signs and modules of the
coefficients which can provide the evolution from a symmetric phase
to the EW symmetry breaking phase with spontaneous mass generation
due to condensation of Higgs fields. Thus one has to inquire about
whether the HD potential has local minimums and what are the
restrictions on the arbitrary coefficients which provide  the
existence of such minimums.

Accordingly we are going to investigate all possible critical
points\footnote{Herein the notion of critical point implies a
stationary one.} of this potential depending on the values of its
coefficients. Without loss of generality one can impose $C>0$. For
the opposite sign of $C$ the set of critical points can be found by
reflection $V\to - V$. One can see, that $V$ has no any critical
points at $H=0$. Let us perform the coordinate transformation to the
variable  $\eta$, \ba H^2 &=& \eta e^{2\phi}\label{13} \ea

Such a transformation is non-degenerate at $H \neq 0$ and preserves
all the information about extremal properties of our potential.

In the new variables the potential takes the form,
\be
V = e^{4\phi}\left(A + B\eta - C \phi \eta^2\right) + E e^{2\phi}
\eta . \label{4}
\ee
Critical point coordinates obey the following equations, \ba
2A + B \eta  - \frac{C}{2} \eta^2   &=& 0 \label{5}\\
\left(\frac{2C\eta}{E}\right)\phi  - \frac{B}{E} &=&  e^{-2\phi}
\label{6} \ea with the additional requirement $\eta > 0$ .

From the equation (\ref{5}) we immediately find,
\be
\eta_{1,2} = {\frac {4A}{-B\pm\sqrt {{B}^{2}+4\,AC}}} .\label{8}
\ee
It is known (for a quick introduction see e.g.~\cite{wiki}), that
the equation of a type $a x + b = p^{c x + d}~~a,c\neq 0$, can be
exactly solved in terms of the Lambert $W(z)$
function~\cite{lambert}. By definition, it is a solution of the
equation,
\be
z = W(z)e^{W(z)} \label{lam}
\ee
The function $W e^W$ is not injective and  $W$ is multivalued
(except for 0). If we look for real-valued $W$ then the relation
\eqref{lam} is defined only for $x \geq 1/e$, and is double-valued
on $(-1/e, 0)$.

Let us introduce the notation $W_{0}(x)$ for the upper branch. It is
defined at $-1/e \leq x < \infty$ and it is monotonously increasing
from -1 to $+\infty$. The lower branch is usually denoted
$W_{-1}(x)$. It is defined only on $-1/e\leq x< 0$ and it is
monotonously decreasing from -1 to $-\infty$.

In these terms the general solution of \eqref{6} is given by,
\be
\phi = \frac12 W \left( \frac{E e^{-\frac{B}{\eta C}}}{\eta C}
\right) +{\frac {B}{2\eta\,C}}
 \label{10}
\ee
Since we have two values of $\eta$ and the real $W$ is
double-valued, then the maximal number of critical points is four.
However $\eta$ must be positive and real, and $\phi$ must be real.
From these requirements one obtains the restrictions on the
coefficients, which provide an existence of each critical point.

We shall denote our critical points as $(m,n)$. Here the first index
$m$ marks the sign $\pm$ and corresponds to the type of a chosen
$\eta$ from (\ref{8}). Index $n$ ranges over $-1,0$ and corresponds
to the chosen branch of $W$ function. We specify a type of each
critical point with the help   of the Hessian matrix eigenvalues and
find the following results for the acceptable composition of
coefficient signs.

We seek for combinations of signs of the coefficients $A,B,C,E$
which  provide a {\it minimum} triggering the spontaneous EW
symmetry breaking at a final stage of cosmic evolution. There are 11
combinations of signs  which are
forbidden as they don't  provide the existence of a local minimum. \\
\begin{center}\begin{tabular}{|c|c|c|c|}\hline
sign (A)&sign(B)& sign(C)&sign(E)\\\hline $\pm$ & $\pm$& +&+\\\hline
-&-&+&-\\\hline -&$\pm$&-&$\pm$\\\hline +&+&-&$\pm$\\\hline
\end{tabular}
\end{center}

Only five combinations of signs can support the required minimum.

\begin{center}\begin{tabular}{|c|c|c|c|}\hline
sign (A)&sign(B)& sign(C)&sign(E)\\\hline + & $+$& +&-\\\hline
+&-&+&-\\\hline -&$+$&+&-\\\hline +&-&-&$+$\\\hline
+&-&-&$-$\\\hline
\end{tabular}
\end{center}

\subsection{Transition from symmetric phase to Electroweak symmetry breaking
phase  and choice of signs}
\begin{figure}
\includegraphics
[scale=.64]{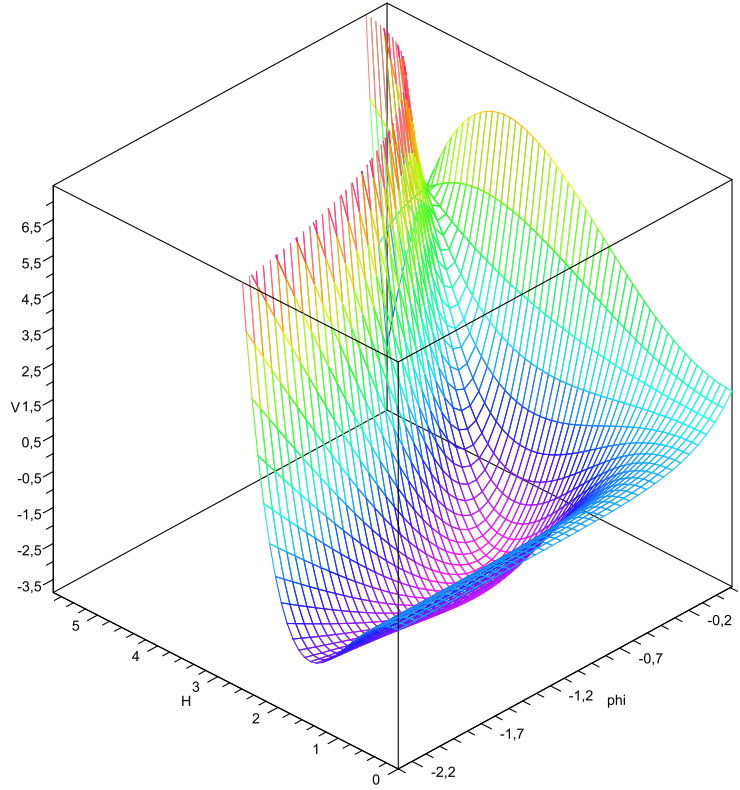}\centering
\caption{View of the effective Higgs-dilaton potential in the vicinity of its two symmetric
local minima: $H =  H_m =  2.29$ and $\phi = \phi_m = -0.72$. Colored  lines
represent the sections of the plot of the potential by the surfaces of constant $\phi$ and constant
$H$.  Parameters are taken as follows: $A = 1$, $B = 2.1$, $C = 0.2$, $E = -2$.} \label{fig1}
\end{figure}

Now let's examine the possibility of
scenario when at the first stage of the Universe evolution one deals
with a massless world with the vanishing v.e.v. of the Higgs field $
\langle H_{in} \rangle = 0$ (symmetric phase).  Thus we adopt that
every initial point $(\phi_{in}, H_{in} = 0)$ for starting evolution
 is acceptable if the function $V|_{\phi = \phi_{in}}(H)$
has a local minimum at $H = 0$ and if we can roll down from the
initial point to the final one which is a local minimum
corresponding to the Higgs phase. We have listed five combinations
of signs of the parameters $A$, $B$, $C$, $E$ which provide the
existence of the local minimum . Nevertheless not all of these
combinations support the above transition scenario. Indeed one can
prove that this scenario can be realized only for positive $A,B,C$
and negative $E$. For this case the solution for minimum belongs to
the class $(+,-1)$ and the minimum (final-stage) coordinates  are
given by, \ba
\eta_{fin} &=& {\frac {4A}{-B+\sqrt {{B}^{2}+4\,AC}}} > 0\label{20},\\
\phi_{fin} &=& \frac12 W_{-1} \left( \frac{E e^{-\frac{B}{\eta_{fin}
C}}}{\eta_{fin} C} \right) +{\frac {B}{2\eta_{fin}\,C}} .\label{21}
\ea The requirement for $\phi$ to be real leads to,
\be
{E_{min} < E < 0,~~~~~E_{min}\equiv -
C\eta_{fin}\exp\left\{ -1 + \frac{B}{\eta_{fin} C} \right\}}
\label{c6}
\ee
The additional bounds exist on the coefficients,
\be
B e^{2\phi_{in}} + E > 0,
\ee
to guarantee that the initial point is in the symmetric phase.
Evidently the phase transition point during evolution appears for
$\phi_{crit} = (1/2) \ln(- E/B) < \phi_{in}$. It can be shown that
$\phi_{fin} <\phi_{crit} <0$ and therefore $B+E > 0$. We remark that
the latter inequality entails $|E_{min}| >|E|$ . We summarize our
finding in Fig. \ref{fig1}.

\section{Conclusions}
The main conclusions of this paper are:
\begin{itemize}
\item {The bosonic spectral action can be provided by restoration of scale invariance in the fermion world or}
\item {the bosonic spectral action can emerge from scale non-invariance of fermion world  in terms of composite dilation; the two effective potentials differ in sign; }
\item {the requirement to trigger EW breaking phase transition during evolution to the Higgs potential minimum gives a favor to the composite nature of dilaton.}
\end{itemize}

\acknowledgments This work has been supported in part by CUR
Generalitat de Catalunya under project 2009SGR502. The work of
A.A.A.\ and M.K.\ was supported by Grant RFBR 09-02-00073 and  SPbSU
grant 11.0.64.2010. M.K.\ is supported by Dynasty Foundation
stipend.


\begin{thebibliography}{99}


\bibitem{Connesbook} A. Connes, \textit{Noncommutative Geometry},
    Academic Press, 1984.


\bibitem{ConnesLott} A.~Connes and J.~Lott,
  ``Particle Models and Noncommutative Geometry (Expanded Version),''
  Nucl.\ Phys.\ Proc.\ Suppl.\  {\bf 18B} (1991) 29.

\bibitem{SpectralAction} A.~H.~Chamseddine and A.~Connes,
  ``The spectral action principle,''
  Commun.\ Math.\ Phys.\  {\bf 186}, 731 (1997)
  [arXiv:hep-th/9606001].


\bibitem{AC2M2} A.~H.~Chamseddine, A.~Connes and M.~Marcolli,
  ``Gravity and the standard model with neutrino mixing,''
  Adv.\ Theor.\ Math.\ Phys.\  {\bf 11} (2007) 991
  [arXiv:hep-th/0610241].




\bibitem{Sakharov} A.~D.~Sakharov,
  ``Vacuum quantum fluctuations in curved space and the theory of
  gravitation,''
  Sov.\ Phys.\ Dokl.\  {\bf 12} (1968) 1040
  [Dokl.\ Akad.\ Nauk Ser.\ Fiz.\  {\bf 177} (1967\ SOPUA,34,394.1991\ GRGVA,32,365-367.2000) 70].


\bibitem{Visser} M.~Visser,
  ``Sakharov's induced gravity: A modern perspective,''
  Mod.\ Phys.\ Lett.\  A {\bf 17} (2002) 977
  [arXiv:gr-qc/0204062].

\bibitem{aano} A.~A.~Andrianov and Yu.~V.~Novozhilov,
  ``Gauge Fields and Correspondence Principle,''
  Theor.\ Math.\ Phys.\  {\bf 67}, 448 (1986)
  [Teor.\ Mat.\ Fiz.\  {\bf 67}, 198 (1986)].


\bibitem{anlizzi} A.~A.~Andrianov and F.~Lizzi,
  ``Bosonic Spectral Action Induced from Anomaly Cancelation,''
  JHEP {\bf 1005} (2010) 057
  [arXiv:1001.2036 [hep-th]].


\bibitem{FellDoran} J.~M.~G.~Fell and R.~S.~Doran, {\it
    Representations of $^*$-Algebras, Locally Compact Groups and
    Banach $^*$-Algebraic Bundles}, Academic Press, (1988).

\bibitem{Ticos} J.M.~Gracia-Bondia, J.C.~Varilly, H.~Figueroa, {\it
    Elements of Noncommutative Geometry}, Birkhauser, 2000.


\bibitem{Landibook} G. Landi, {\it  An Introduction to
    Noncommutative Spaces and their Geometries}, {\sl Springer
    Lecture Notes in Physics 51}, Springer Verlag (Berlin
    Heidelberg) 1997. arXiv:hep-th/9701078.

\bibitem{Madore} J.~Madore,
  ``An Introduction To Noncommutative Differential Geometry and its Physical
  Applications,''
  Lond.\ Math.\ Soc.\ Lect.\ Note Ser.\  {\bf 257} (2000) 1.

\bibitem{Madoreearly} J.~Madore,
  ``Kaluza-Klein Aspects Of Noncommutative Geometry,''
{\it  In *Chester 1988, Proceedings, Differential geometric methods
in theoretical physics* 243-252.}

\bibitem{D-VKM} M.~Dubois-Violette, R.~Kerner and J.~Madore,
  ``Noncommutative Differential Geometry Of Matrix Algebras,''
  J.\ Math.\ Phys.\  {\bf 31} (1990) 316.


\bibitem{Fujikawabook} K.~Fujikawa, H.~Suzuki, {\it Path
    Integrals And Quantum Anomalies}, Oxford university Press, 2004.

\bibitem{ChamseddineConnesscale}
  A.~H.~Chamseddine and A.~Connes,
  ``Scale invariance in the spectral action,''
  J.\ Math.\ Phys.\  {\bf 47} (2006) 063504
  [arXiv:hep-th/0512169].



\bibitem{LMMS} F.~Lizzi, G.~Mangano, G.~Miele and G.~Sparano,
  ``Fermion Hilbert space and fermion doubling in the noncommutative  geometry
  approach to gauge theories,''
  Phys.\ Rev.\  D {\bf 55}, 6357 (1997)
  [arXiv:hep-th/9610035].


\bibitem{G-BIS} J.~M.~Gracia-Bondia, B.~Iochum
    and T.~Schucker,
  ``The standard model in noncommutative geometry and fermion doubling,''
  Phys.\ Lett.\  B {\bf 416}, 123 (1998)
  [arXiv:hep-th/9709145].

  \bibitem{AndrianovBonoraGamboa}
  A.~A.~Andrianov, L.~Bonora and R.~Gamboa-Saravi,
  ``Regularized Functional Integral For Fermions And Anomalies,''
  Phys.\ Rev.\  D {\bf 26}, 2821 (1982).

\bibitem{AndrianovBonora1}
  A.~A.~Andrianov and L.~Bonora,
  ``Finite - Mode Regularization Of The Fermion Functional Integral,''
  Nucl.\ Phys.\  B {\bf 233}, 232 (1984).

\bibitem{AndrianovBonora2}
  A.~A.~Andrianov and L.~Bonora,
  ``Finite Mode Regularization Of The Fermion Functional Integral. 2,''
  Nucl.\ Phys.\  B {\bf 233}, 247 (1984).

  \bibitem{AANN}
 A.~A.~Andrianov, V.~A.~Andrianov, V.~Y.~Novozhilov and Yu.~V.~Novozhilov,
  ``Joint Chiral and Conformal Bosonization in QCD and the Linear Sigma
  Model,''
  Phys.\ Lett.\  B {\bf 186} (1987) 401.

\bibitem{NovozhilovVassilevich} Y.~V.~Novozhilov and
    D.~V.~Vassilevich,
  ``Induced Quantum Conformal Gravity,''
  Phys.\ Lett.\  B {\bf 220} (1989) 36.


\bibitem{Sitarz}
  A.~Sitarz,
  ``Spectral action and neutrino mass,''
  Europhys.\ Lett.\  {\bf 86} (2009) 10007
  [arXiv:0808.4127 [math-ph]].

\bibitem{Vassilevich:2003xt}
  D.~V.~Vassilevich,
  ``Heat kernel expansion: User's manual,''
  Phys.\ Rept.\  {\bf 388}, 279 (2003)
  [arXiv:hep-th/0306138].

  \bibitem{NelsonSakellariadu}
  W.~Nelson and M.~Sakellariadou,
  ``Cosmology and the Noncommutative approach to the Standard Model,''
  arXiv:0812.1657 [hep-th].


\bibitem{MarcolliPierpaoli} M.~Marcolli and E.~Pierpaoli,
  ``Early Universe models from Noncommutative Geometry,''
  arXiv:0908.3683 [hep-th].

\bibitem{wiki} http://en.wikipedia.org/wiki/Lambert$\_$function .

\bibitem{lambert} Lambert, J. H. "Observations variae in Mathesin
    Puram." Acta Helvitica,
    physico-mathematico-anatomico-botanico-medica 3, 128-168, 1758.
    Euler, L. "De serie Lambertina Plurimisque eius insignibus
    proprietatibus." Acta Acad. Scient. Petropol. 2, 29-51, 1783.

\bibitem{shaposh} F.~Bezrukov, A.~Magnin, M.~Shaposhnikov and
    S.~Sibiryakov,
  ``Higgs inflation: consistency and generalisations,''
  JHEP {\bf 1101}, 016 (2011)
  [arXiv:1008.5157 [hep-ph]].


\end{thebibliography}
\end{document}